# Bi-directional Alfvén Cyclotron Instabilities in the Mega-Amp Spherical Tokamak

S.E. Sharapov, M.K.Lilley[1], R.Akers, N. Ben Ayed, M.Cecconello[2], J.W.C.Cook[3], G.Cunningham, E.Verwichte[3] and the MAST Team

*CCFE, Culham Science Centre, Abingdon OX14 3DB, UK*
[1]*Physics Department, Imperial College, London, SW7 2AZ, UK*
[2]*Department of Physics and Astronomy, Uppsala University, SE-75105 Uppsala, Sweden*
[3] *Centre for Fusion, Space and Astrophysics, Department of Physics, University of Warwick, Coventry CV4 7AL, UK*

**e-mail** contact of main author: Sergei.Sharapov@ccfe.ac.uk

Alfvén cyclotron instabilities excited by velocity gradients of energetic beam ions were investigated in MAST experiments with super-Alfvénic NBI over a wide range of toroidal magnetic fields from ~0.34 T to ~0.585 T. In MAST discharges with high magnetic field, a discrete spectrum of modes in the sub-cyclotron frequency range is excited toroidally propagating counter to the beam and plasma current (toroidal mode numbers $n < 0$). At lower magnetic field $\leq 0.45$ T, a discrete spectrum of Compressional Alfvén Eigenmodes (CAEs) with $n > 0$ arises, in addition to the modes with $n < 0$. At lowest magnetic fields, the CAEs with $n > 0$ become dominant, they are observed in frequency range from ~250 kHz for $n = 1$ to ~3.5 MHz for $n = 15$, well above the on-axis ion cyclotron frequency (~2.5 MHz). The data is interpreted in terms of normal and anomalous Doppler resonances modified by magnetic drift terms due to inhomogeneity and curvature of the magnetic field. A Hall MHD model is applied for computing the eigenfrequencies and the spatial mode structure of CAEs and a good agreement with the experimental frequencies is found.

## 1. Introduction

Energetic particles may excite Alfvénic instabilities due to the sources of free energy caused by: 1) radial gradient of the energetic (beam) particle pressure, $\nabla p_b \neq 0$, 2) bump-on-tail of the beam distribution function $F_b$ in energy, $\partial F_b / \partial E > 0$, and 3) temperature anisotropy of the beam, $\partial F_b / \partial v_\perp^2 \neq \partial F_b / \partial v_\parallel^2$, where $v_\perp$, $v_\parallel$ are velocity components perpendicular and parallel to the magnetic field [1]. The instabilities driven by $\nabla p_b$, such as Toroidal Alfvén Eigenmodes (TAE) and fishbones, cause a radial re-distribution/ loss of energetic ions, while the instabilities driven by the velocity gradients may cause a beam energy loss and pitch-angle scattering of the beam ions higher than the classical Coulomb collisions. Instabilities in the ion-cyclotron range of frequency driven by cyclotron resonance interaction with energetic ions has been long recognised as an important channel of releasing the free energy sources due to the velocity gradients [1, 2], and a first theory of weakly-damped eigenmodes in the ion-cyclotron frequency range in realistic plasma configurations was developed in the 1980$^{th}$ [3, 4].

Ion Cyclotron Emission [5] observed in the earliest JET deuterium-tritium (DT) plasmas has set the stage for a systematic experimental study of the velocity gradient-driven cyclotron instabilities on tokamaks [6, 7]. On JET, the ICE phenomenon was interpreted as a cyclotron instability of compressional Alfvén (also called fast Alfvén or magneto-acoustic waves) eigenmodes (CAEs) excited at alpha-particle harmonics, $\omega \cong l\omega_{B\alpha}$, by centrally born, marginally trapped alpha-particles undergoing drift excursions to the plasma edge where CAEs are localised. Here $\omega_{B\alpha}$ is the cyclotron frequency of alpha-particles and $l$ is an integer number. Since only highly energetic trapped alpha-particles have orbits large enough to reach the edge plasma region, the velocity distribution of the alpha particles in the CAE region has a local bump-on tail, $\partial F_\alpha / \partial E > 0$, at a certain speed and pitch angle. A comprehensive theory of ICE has been developed (see, e.g. [8,9]) convincingly interpreting the ICE data. In particular, the experimentally observed fine splitting of the mode frequencies was explained in [10] as a result of the toroidal drift of trapped energetic ions, which modifies the resonance as $\omega \cong l\omega_{B\alpha} \pm mV_{D\alpha}/r$, where $V_{D\alpha}$ is the toroidal drift velocity of trapped fast ions, and the poloidal mode number, $m > 0$, and $m < 0$, provide two maxima to the instability growth rate and cause the frequency splitting. During the high fusion power DT campaign on JET, further study of ICE has shown that the intensity of the spectral ICE lines was proportional to the DT fusion rate thus confirming the alpha-particle drive as the source of CAEs excitation [11]. This observation opens the opportunity for using the ICE measurements as a passive diagnostic of the alpha-particles in burning DT plasmas on ITER.

Development of the Spherical Tokamak (ST) concept and the two new large ST machines built, NSTX [12] and MAST [13] with super-Alfvénic NBI, have significantly expanded the opportunities for experimental studies of energetic particle-driven Alfvén instabilities, including those in the frequency range comparable to ion cyclotron frequency [14, 15]. Discrete spectra of modes with frequencies below ion cyclotron frequency driven by nearly tangentional NBI were first observed on NSTX [14], and later on MAST [15]. These modes were identified as CAEs and Global Alfvén Eigenmodes (GAEs) from the Alfvénic character of their frequency evolution with plasma density and magnetic field, the mode polarisation, and through a comparison with theoretically derived spectra of CAE and GAE in the ST geometry [16, 17]. In contrast to the ICE case associated with edge-localised CAEs, the modes driven by NBI in STs were found to be at mid-radius [15]. Although the free energy source of the instability in NSTX [14] was found to be similar to that of ICE, $\partial F_b / \partial E > 0$, the sub-cyclotron range of the mode frequencies and the dominant component of the beam velocity parallel to the magnetic field, required a Doppler shifted wave-particle resonance (1) to be relevant. Indeed, one of the main features of the beam-driven modes in STs is that, in comparison to the ICE data, the mode frequencies deviate significantly from integer multiples of ion cyclotron frequency. Importance of this for the sub-cyclotron frequency modes can be seen from the resonance condition

$$\omega = k_\parallel v_{\parallel b} + l\omega_{Bb} + \vec{k}_\perp \cdot \vec{V}_{Db}, \tag{1}$$

where $k_\parallel \equiv \vec{k} \cdot \vec{B} / B$ and $\vec{k}_\perp = \vec{k} \times \vec{B} / B$ are the wave-vectors parallel and perpendicular to the magnetic field, and

$$\vec{V}_{Db} = \frac{v_\perp^2 / 2 + v_\parallel^2}{\omega_{Bb}} \left[ \mathbf{b} \times \frac{\nabla B}{B} \right] \tag{2}$$

is the magnetic drift velocity of the beam ions. Doppler frequency shifts well below $\omega = l\omega_{Bb}$ occur if $k_\parallel v_{\parallel b} + \vec{k}_\perp \cdot \vec{V}_{Db} \approx -l\omega_{Bb}$, and since the nearly tangentional NBI generates beam with $v_{\parallel b} \gg V_{Db}$, the parallel wave-vector becomes an important parameter determining the Doppler shift. It is easy to see next that when $\left| k_\parallel v_{\parallel b} \right| / \left| \vec{k}_\perp \cdot \vec{V}_{Db} \right| \gg 1$, the normal Doppler resonance with $l = +1$ gives the frequency range below ion cyclotron

frequency if $k_\| v_{\|b} < 0$ thus determining the directivity of the wave propagation with respect to the parallel beam velocity.

The directivity of the mode propagation excited in the sub-cyclotron frequency range was studied in detail on MAST [18, 19]. It was found that i) with rare exceptions, almost all the modes propagate counter to the beam and the plasma current, i.e. they have negative toroidal mode number, $n < 0$, and ii) modes with higher frequencies have lower mode numbers $|n|$. It became interesting then to search for some MAST reproducible plasma scenarios, in which the beam would excite waves propagating co-beam and co-current, $k_\| v_{\|b} > 0$. Anomalous Doppler resonance for $l = -1$ becomes relevant then,

$$\omega = k_\| v_{\|b} - \omega_{Bb} + \vec{k}_\perp \cdot \vec{V}_{Db}, \tag{3}$$

and the free energy source associated with the beam temperature anisotropy, $T_{\|b} > T_{\perp b}$, plays a role [20,1]. Although the resonance (3) is well-known for causing a fan-type (i.e. with a strong pitch-angle scattering) instability of run-away electron beams in tokamaks [21], similar effects for ion beams were not explored in detail yet and are of great interest. However, in order to access this particular resonance in infinite homogeneous plasma (i.e. in the absence of the drift term), a threshold with respect to Alfvén velocity $V_A$ of $v_{\|b}/V_A \approx 2.6$ needs to be reached. Although toroidal geometry causing the drift in (3) may soften the requirement, still the best way of approaching the anomalous Doppler resonance driving CAEs is to decrease the equilibrium magnetic field and thereby obtain the lowest possible values of Alfvén velocity (at constant beam velocity).

We present results of this low-field experiment on MAST in Section 2. Numerical calculations of eigenmodes using the WHALES Hall-MHD model are presented in Section 3. Section 4 discusses the resonance condition, and Section 5 presents conclusions.

## 2. Transitions in the spectra of Alfvénic modes at decreasing magnetic field

MAST is a low aspect ratio tokamak with typical major and minor radii of $R_0 = 0.86$ m and $a = 0.6$ m respectively. Looking from above the machine, the equilibrium toroidal field is in the clockwise direction, while the inductive plasma current and neutral beam injection (NBI) are in the anti-clockwise direction. The tangency radius of NBI is 0.7 m, and in the experiment deuterium (D) NBI with maximum energy $E_b^{max} \approx 65$ keV was injected into D plasma thus generating a D beam distribution function with maximum velocity of $V_b^{max} \approx 2.5 \cdot 10^6$ m/s. The range of operational parameters used for the experiment discussed here was as follows: toroidal field at the magnetic axis $B_T \approx 0.34 - 0.585$ T, maximum plasma current $I_P^{max} \approx 600$ kA, electron density in the plasma center $n_e(0) \approx (2-4) \cdot 10^{19}$ m$^{-3}$, and electron temperature in the plasma center $T_e \approx (0.8 - 1)$ keV. The ratio of beam velocity to Alfvén velocity was varied in the range $V_b^{max}/V_A = 1.54 - 2.26$. The best CAE data was collected when NBI power of ~2 MW was used, while doubling of NBI power caused significant activity of energetic particle-driven modes in the TAE and fishbone frequency ranges, which dominated the spectrum of the measured Alfvénic perturbations. During the experiment, MAST discharges were in L-mode most of the time, with some short-time transitions into H-mode. Figure 1 shows toroidal magnetic field $B_T$, inductive current $I_P$, and NBI power waveforms in three MAST discharges with three different magnetic fields. For detecting electromagnetic waves, 10 OMAHA coils digitised to 10 MHz sampling rate were used allowing measurements up to 5 MHz to be performed [4]. For determining electron density and electron temperature profiles, Thomson scattering diagnostics were employed with high spatial resolution, and a motional Stark effect (MSE) diagnostic was used for measuring the safety factor profiles.

In the experiment, $B_T$ was varied from discharge to discharge in order to cover the whole range of toroidal magnetic fields possible on MAST. As $B_T$ decreased, significant changes were observed in both the frequency spectrum and toroidal mode numbers $n$ of modes with frequencies in the ion cyclotron range as shown in Figures 3-5.

At high $B_T \approx 0.5$ T, only modes propagating counter-beam, i.e. with $n < 0$ are excited. Figure 3 shows both amplitude and phase magnetic spectrograms of the modes excited. The frequency range of the modes observed lies in the range from ~700 kHz to ~1.4 MHz, while the on-axis cyclotron frequency is $f_{Bi}(0) \equiv \omega_{Bi}(0)/(2\pi) \approx 3.8$ MHz so that $f \equiv \omega/(2\pi) \approx (0.18 \div 0.37) f_{Bi}(0)$ (Fig.3). The phase magnetic spectrogram shows toroidal mode numbers of the modes excited, $n = -5, ..., -10$.

At somewhat lower magnetic field, $B_T \sim 0.4$ T (MAST discharge #27145), in addition to the modes in the sub-cyclotron frequency range ~700 kHz – 1.2 MHz, another type of discrete spectrum arises in a much higher frequency range, $f \approx 1.8$ MHz – 2.2 MHz as

Figure 4 shows. This frequency range is closer to the on-axis cyclotron frequency of ~3 MHz and is comparable to the ion cyclotron frequency at the outer edge of the plasma, ~1.9 MHz. A zoom of the phase magnetic spectrogram shows that this spectrum has positive toroidal mode numbers $n = 8, 9, 10$, much wider gaps between the frequencies of the modes with different $n$'s, $f_{n+1} - f_n \approx 220$ kHz, and frequencies of the modes increasing with toroidal mode number. For these machine and plasma parameters, the ratio between the beam velocity and Alfvén velocity was $V_b^{max}/V_A \approx 2$.

Finally, the MAST discharge in this experiment with the lowest magnetic field, $B_T \sim 0.34$ T, exhibited massive activity of modes in the range from ~250 kHz to ~3.5 MHz, with toroidal mode numbers $n = 1, ..., 15$. Figure 5 shows the spectrogram and mode numbers for this discharge, #27148. In this case, the modes with $n < 0$ in the sub-cyclotron frequency range ceased to exist, the gaps between the frequencies of the modes with different $n$'s become $f_{n+1} - f_n \approx 250$ kHz and modes with higher $n$'s have higher frequencies. For this discharge, the ratio between the beam velocity and Alfvén velocity was $V_b^{max}/V_A \approx 2.26$. The modes of the highest frequency range exceed the on-axis cyclotron frequency of ~2.6 MHz, so the modes excited are identified as CAE. The measured profiles of electron temperature and density around the time of CAE observation at 100 ms are shown in Figure 6.

During the experiment, a neutron camera was employed to measure the profiles of DD neutrons, which were mostly produced by beam-plasma reactions [22]. Although some significant drops in the DD neutron rates were observed in some of the discharges, no obvious effects of the CAEs on the neutron rates were observed.

In the discharge with lowest $B_T$, the discrete spectrum of CAEs extended down to the extreme value of $n = 1$, observed at the very low frequency of ~250 kHz (see Fig.5). A similar $n = 1$ mode was observed in another MAST discharge #27147 with $B_T \sim 0.38$ T as Figure 7 shows. In this case, no low-frequency long-lasting modes or any other significant MHD activity was seen at the times of CAE excitation, e.g. ~100 ms. This case was taken for the further modelling of the CAEs presented in the next Section.

Finally, CAEs in MAST discharge #27148 seem to exhibit some fine structure of the frequency spectrum. Such fine structure was also seen more clearly in a number of other discharges, e.g. in MAST pulse #30080 shown in Figure 8. The frequency split between the modes with the same toroidal mode numbers, is ~40 kHz, which is much smaller than the frequency separation between CAEs with different $n$'s (about 250 kHz), but higher than

Doppler shift caused by the beam-driven toroidal rotation of the plasma, ~ 15 kHz. Such fine splitting may be caused by different poloidal mode numbers, similar to the fine splitting observed in ICE [10], or may result from nonlinear effects in the wave-particle interaction similar to the pitchfork splitting of TAE [23].

### 3. Modelling the spectra of Compressional Alfvén Eigenmodes with Hall MHD

For modelling Compressional waves, a general Hall-MHD finite element spectral code called WHALES (Warwick Hall-MHD arbitrary linear eigenvalue solver) is employed, which builds on the experience of [24]. In the model used, the Hall term is incorporated by using the modified plasma displacement

$$\vec{\eta} = \vec{\xi} - \frac{i\omega}{\omega_{Bi}} \frac{\vec{\xi} \times \vec{B}_0}{B_0}, \tag{4}$$

where $\vec{B}_0$ is the equilibrium magnetic field and $\vec{\xi} = -(c/i\omega B_0^2)[\delta\vec{E} \times \vec{B}_0]$ is the plasma displacement due to the cross-field drift velocity associated with perturbed electric field $\delta\vec{E}$. The linearized induction equation then reduces to $\delta\vec{B} = \vec{\nabla} \times [\vec{\eta} \times \vec{B}_0]$. The components of the modified plasma displacement perpendicular to $\vec{B}_0$, which are to be determined, have the form:

$$\eta_\psi = \vec{\eta} \cdot \vec{\nabla}\psi, \tag{5}$$

$$\eta_\Lambda = \vec{\eta} \cdot \frac{[\vec{B}_0 \times \vec{\nabla}\psi]}{B_0^2}, \tag{6}$$

and a right-handed coordinate system $(\nabla\psi, \nabla\phi, \nabla\vartheta)$ is employed, with toroidal angle $\phi$ increasing anti-clockwise when viewed from above. In contrast to the model developed in [24], which solved equations for three magnetic field components, now the equations for only two displacement variables are solved.

In order to obtain a well-converged spectrum of compressional Alfvén waves with $k_\parallel /|k| < 1$, terms linear in the parallel wave-vector are retained, while terms $\propto (k_\parallel /|k|)^2$ are omitted. In this way, the problem of the coupling of the compressional Alfvén and shear Alfvén waves is resolved in the model used.

Compressional Alfvén Eigenmodes can have global mode structure, i.e. radial widths comparable to the minor radius, so it is necessary to have good spatial resolution in the whole poloidal cross-section. A finite element method is used in the radial direction, with a spectral method in poloidal direction, and a single harmonic in the toroidal direction. Solutions of the Hall MHD equations are sought as sums of hybrid Fourier and finite elements

$$\eta_j = \sum_{l=1}^{L} \sum_{m=M_{\min}}^{M_{\max}} c_{l,m} \Lambda_l(\psi) e^{im\vartheta} e^{in\phi}, \tag{7}$$

where $\Lambda_l(\psi)$ is the $l^{th}$ element in the $\nabla\psi$ direction, and $c_{l,m}$ is the complex amplitude of the hybrid element of the displacement. A large number of poloidal harmonics and 5th order b-splines are used as the conforming finite elements for both variables, (5), (6) when the Hall term is large. If the Hall term is small, the variable $\eta_\psi$ is represented with 5th order b-splines, and $\eta_\Lambda$ - with 4th order b-splines. In solving the Hall MHD equations no assumption is made with regard to the self-adjointness of the system, so in general the linearly stable Hall-MHD modes have non-zero imaginary components.

For the Hall-MHD modelling, experimental data was taken from MAST discharge #27147 at $t = 100$ ms. At that time, CAE modes were excited with toroidal mode numbers from $n = 1$ to $n = 15$ as Figure 6 shows. Figure 9 shows the safety factor profile (at $Z = 0$) used in the modelling. Several eigenmodes with somewhat different mode structure and eigenfrequency were computed for every toroidal mode number, with typical well-converged solutions for $n = 1$ and $n = 10$ shown in Figures 10 and 11. The frequency separation between the modes was found to be close to 40 kHz possibly explaining the experimentally observed fine structure of the CAE spectrum.

## 4. Effects of the magnetic field inhomogeneity and curvature on the cyclotron resonances

The magnetic drift of energetic trapped ions was shown to play an important role in the theory of ICE [10], and one can expect that such effects cannot be neglected for the case of passing beam ions. Indeed, considering the resonance condition (3) for decreasing magnetic fields, one notes that the magnetic drift term $V_{Db} \propto 1/\omega_{Bb}$ increases while the cyclotron frequency term itself decreases. A theory describing the effects of passing ion drifts on cyclotron instabilities was developed in [25]. It was shown there for the simple case of a large aspect ratio axisymmetric tokamak with circular cross-section that the drift velocity term for passing fast ions gives a modified cyclotron resonance condition, which in the case of anomalous Doppler resonance can be written as

$$\omega = \left(k_\parallel + \frac{s}{qR}\right)v_{\parallel b} - \omega_{Bb}, \quad s\text{ - integer}. \tag{8}$$

The presence of the $s$ in this expression results from the poloidal dependence of the cyclotron frequency, $\omega_{Bb} = \omega_{Bb}(0) \cdot (1 - (r/R)\cos\vartheta)$, and the magnetic drift velocity, with components $V_{Dr} \propto \sin\vartheta$, $V_{D\vartheta} \propto \cos\vartheta$. Although the resonances with non-zero $s$ provide a smaller particle-to-wave power transfer than the $s = 0$ resonance, the very existence of such resonances makes the critical beam velocity condition described in Section 1, less restrictive. Comparing now the second and third terms in (8), we find that these terms cancel if

$$s \approx qR\omega_{Bb}/v_{\parallel b} = q(R/\rho_{\parallel b}), \tag{9}$$

which gives $s$ as low as 5-6 when $q \approx 1$ is considered as in Fig. 9 and $\rho_{\parallel b}^{\max} = V_b/\omega_{Bb} \sim 15$ cm is taken at $B_T \sim 0.34$ T. It is then possible for the drift-modified anomalous Doppler resonance condition to be satisfied (and thus $n > 0$ CAEs to be excited) for $\omega \ll \omega_{Bb}$, as observed experimentally in MAST discharges #27147 and #27148. This is illustrated schematically in Figure 12. The dispersion curves of shear and compressional Alfvén waves are labelled respectively as SAW and CAW, and the red dashed lines indicate the ion cyclotron frequency and the reference lines $\omega = \pm k_\parallel V_A$. The two black lines show the anomalous Doppler resonance (without the drift term) for $v_{\parallel b} < v_\parallel^{crit}$ and $v_{\parallel b} = v_\parallel^{crit}$, where $v_\parallel^{crit}$ is the threshold velocity for this resonance to occur for the CAW, i.e. the velocity such that the resonance line is tangential to the CAW dispersion curve. As shown in the Figure, this typically occurs at frequencies of the order of $\omega_{Bb}$. However when the drift resonances are taken into account (the blue lines in Figure 12), it can be seen that the crossing points between the resonance lines and CAW dispersion curve is possible at much lower frequencies.

Figure 13 shows schematically the relative location of the drify-modified anomalous Doppler resonances with respect to the beam distribution function on MAST. The thick black dot represents the point in velocity space at which most energetic beam ions are injected (at a particular spatial location in the plasma). The vertical solid line to the right of this represents the parallel velocity corresponding to the anomalous Doppler resonance, unmodified by the drift effects, when the threshold velocity is above the injection velocity. The two dashed lines show the parallel velocities corresponding to the drift-modified anomalous Doppler resonances for $s=1$ and $s=2$. These resonances occur in regions of velocity space where energetic ions are present, and the distribution function is highly anisotropic and thus capable in principle of driving CAEs. We infer that the threshold for CAE excitation via drift-modified anomalous Doppler resonance in toroidal geometry is somewhat lower than the value of 2.6 $V_A$ discussed in Section 1.

## 5. Conclusions

A dedicated MAST experiment with decreasing magnetic field at fixed beam energy has shown the excitation of bi-directional Alfvén cyclotron instabilities. At higher fields exceeding ~0.45 T at the magnetic axis, the excitation of eigenmodes with sub-cyclotron frequencies and mostly negative toroidal mode numbers was observed, while at magnetic fields ≤0.45 T eigenmodes with $n>0$ start to emerge in the frequency range close to, or exceeding, the ion cyclotron frequency. These modes with $n>0$ completely dominate the spectrum of the Alfvén cyclotron instabilities at the lowest magnetic fields ~0.34 – 0.38 T. It is concluded that the modes with $n>0$ are CAEs, due to the frequencies of high-$n$ modes exceeding the ion cyclotron frequency in MAST.

Modelling of the CAEs was performed with the numerical Hall-MHD code WHALES and it was found that this model gives a good agreement with experimentally measured frequencies.

An interpretation of the transition from dominant $n<0$ mode excitation to dominant $n>0$ mode excitation has been proposed in terms of the anomalous Doppler resonance, and it has been shown that magnetic drift terms play a significant role, as in the case of normal Doppler resonance considered earlier. Specifically, drift effects provide a credible explanation for the observed excitation of CAEs with $n>0$ at frequencies well below the ion cyclotron frequency.

## Acknowledgment

This work was part-funded by the RCUK Energy Programme [grant number EP/I501045] and by the European Union's Horizon 2020 research and innovation programme. To obtain further information on the data and models underlying this paper please contact PublicationsManager@ccfe.ac.uk. The views and opinions expressed herein do not necessarily reflect those of the European Commission. The authors thank K.G.McClements for support and encouragement in preparing this paper.

**Figure captions**

Fig.1. Time traces of vacuum equilibrium magnetic field, plasma current, and NBI power waveform in MAST discharges #27143 (highest $B_T$), #27145, and #27148 (lowest $B_T$).

Fig.2. Zoom showing $B_T(0)$ temporal evolution in the comparison discharges shown in Figure 1.

Fig.3. Top: Magnetic spectrogram showing amplitude of $\delta B$ perturbations excited in sub-cyclotron frequency range in MAST discharge #27143 with high magnetic field $B_T \approx 0.5$ T; Bottom: Phase magnetic spectrogram showing toroidal mode numbers of the modes excited.

Fig.4. (top) Magnetic spectrogram showing amplitude of $\delta B$ perturbations excited in MAST discharge #27145 with magnetic field $B_T \approx 0.4$ T; (bottom) Zoom of the phase magnetic spectrogram showing toroidal mode numbers of the higher frequency modes.

Fig.5. (top) Magnetic spectrogram showing amplitude of $\delta B$ perturbations excited in MAST discharge #27148 with magnetic field $B_T \approx 0.34$ T; (bottom) Phase magnetic spectrogram showing toroidal mode numbers of the modes excited.

Fig.6. Electron density and electron temperature profiles measured with high spatial resolution by Thomson scattering diagnostics. Top: $n_e(R)$-profile; middle: $T_e(R)$-profile, and bottom: time slices when the measurements were taken.

Fig.7. (top) Magnetic spectrogram showing amplitude of $\delta B$ perturbations excited in frequency range ~250 kHz – 3.5 MHz in MAST discharge #27147 with magnetic field $B_T \approx 0.34$ T; (bottom) The phase magnetic spectrogram showing positive toroidal mode numbers of the modes, $n = 1,\ 9,\ 15$.

Fig.8. Top: CAEs observed in MAST discharge #30080 ($B_T = 0.37$ T, $E_b = 69$ keV). Bottom: Zoom showing toroidal mode numbers for all modes with fine splitting of the frequency.

Fig.9. Safety factor profile reconstructed with the use of MSE for MAST discharge # 27147 at t=100 ms.

Fig.10. WHALES modelling showing the $n=1$ CAE structure for $\eta_\psi$ (upper plots) and $\eta_\Lambda$ (lower plots) computed for MAST equilibrium # 27147 at t=100 ms. The mode eigenfrequency is Re $f$ = 196.5 kHz, Im $f$ = 14.5 kHz.

Fig.11. WHALES modelling showing the $n=10$ CAE structure $\eta_\psi$ (upper plots) and $\eta_\Lambda$ (lower plots) computed for MAST equilibrium # 27147 at t=100 ms. The mode eigenfrequency is Re $f$ = 2.23 MHz, Im $f$ = 288 Hz.

Fig.12. Schematic plot of compressional Alfvén wave (CAW) and shear Alfvén wave (SAW) dispersion relations with superimposed anomalous Doppler resonances (black lines) and anomalous Doppler resonances modified by the inhomogeneity and curvature of the magnetic field (blue lines).

Fig.13. Schematic plot showing the positions of anomalous Doppler resonances (modified by the magnetic drift), with respect to the beam ion distribution $F_b(v_\perp, v_\parallel)$ in the low-$B_T$ case.

Figure 1.

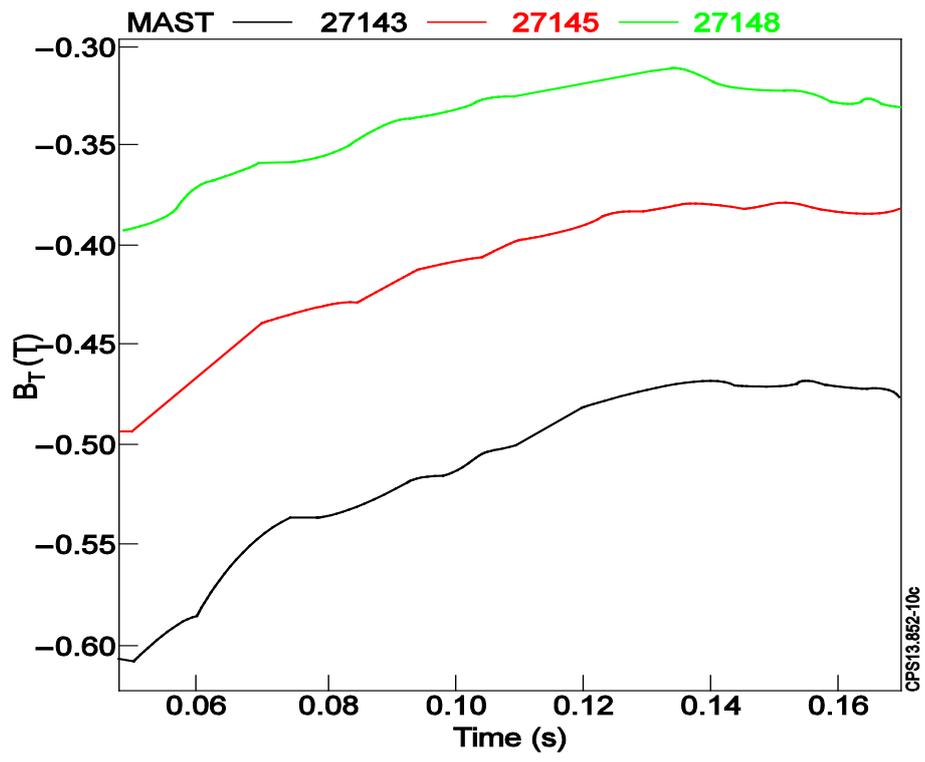

Figure 2.

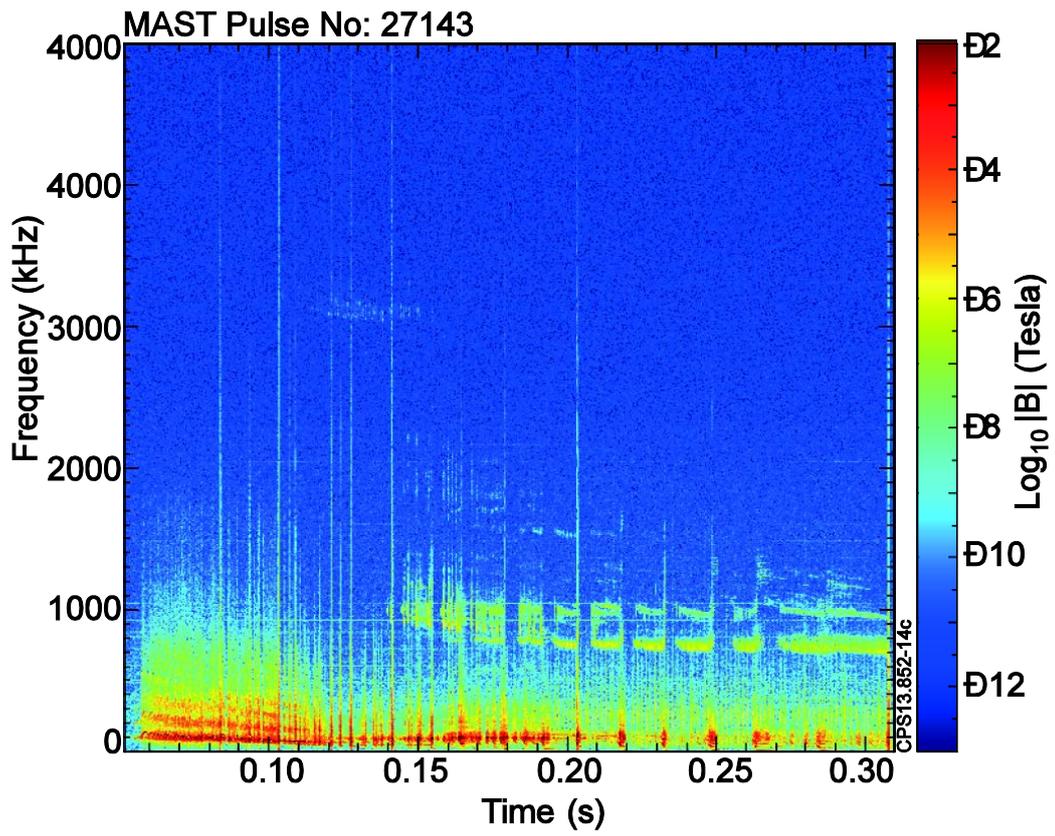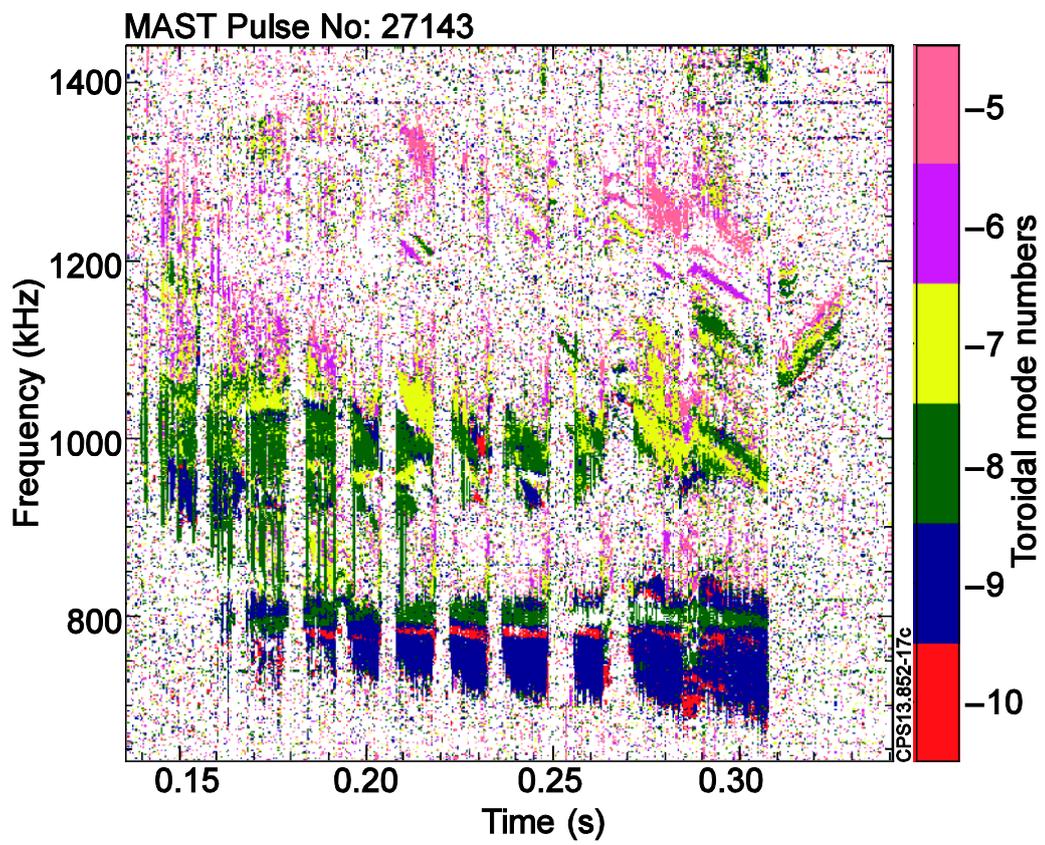

Figure 3.

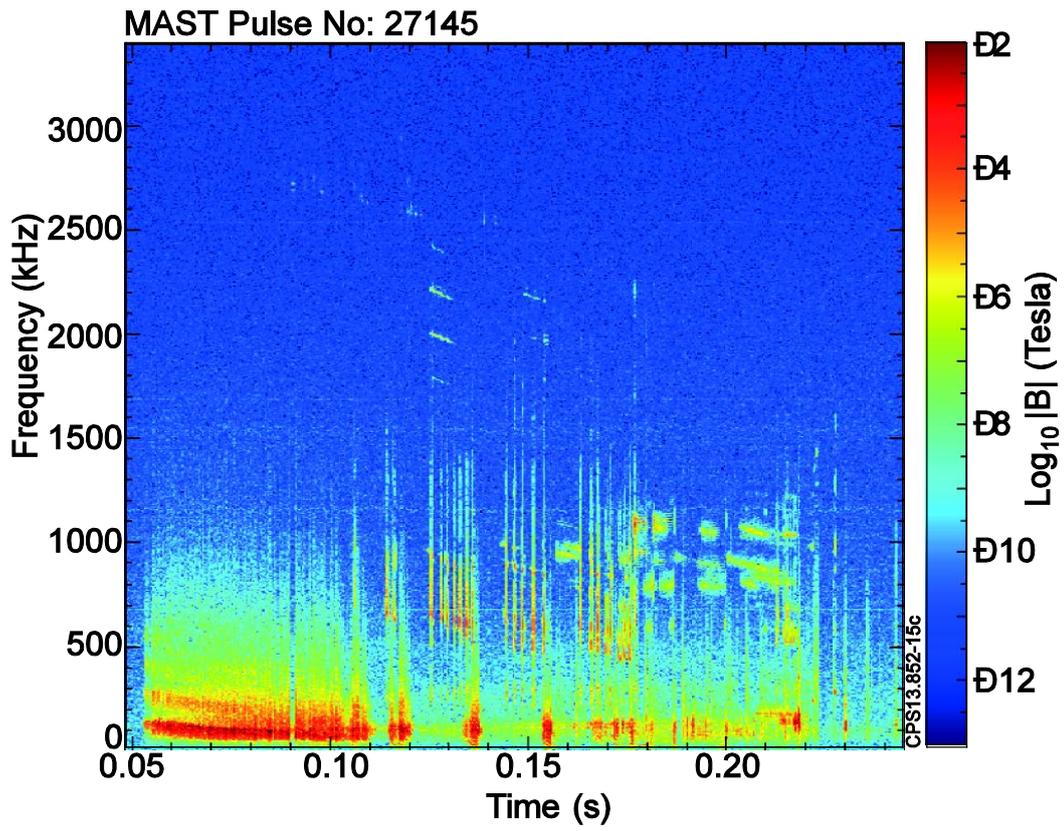

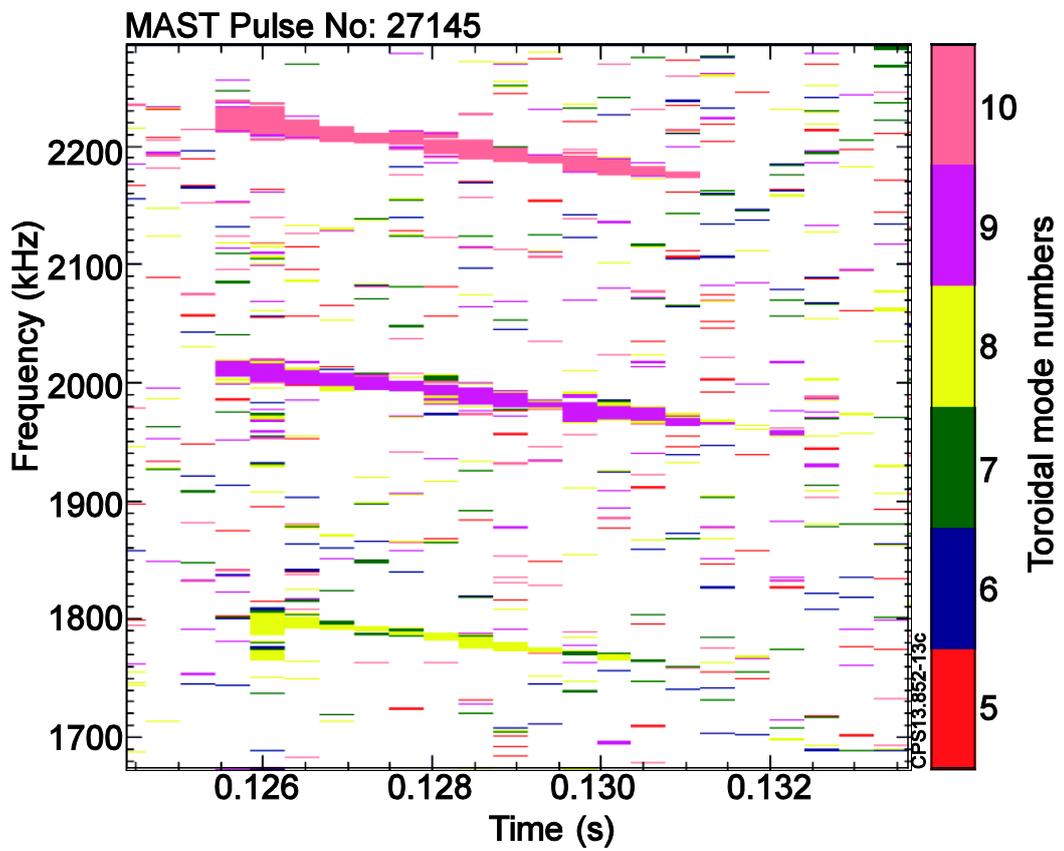

Figure 4.

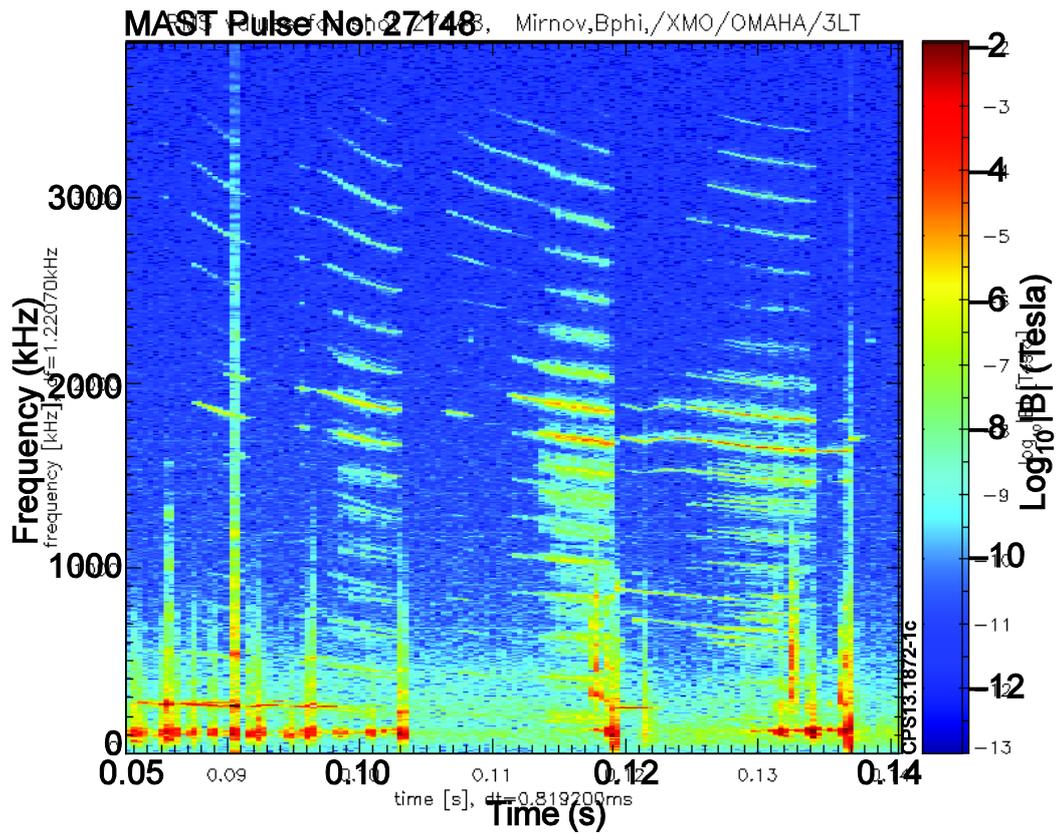

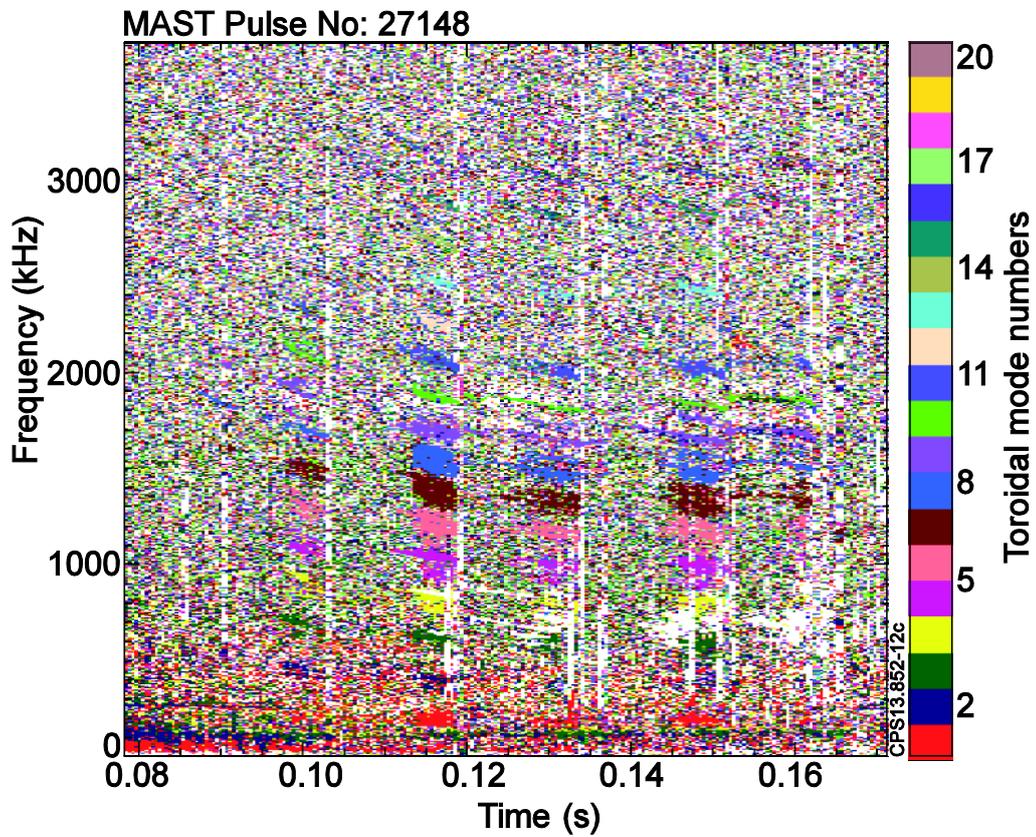

Figure 5.

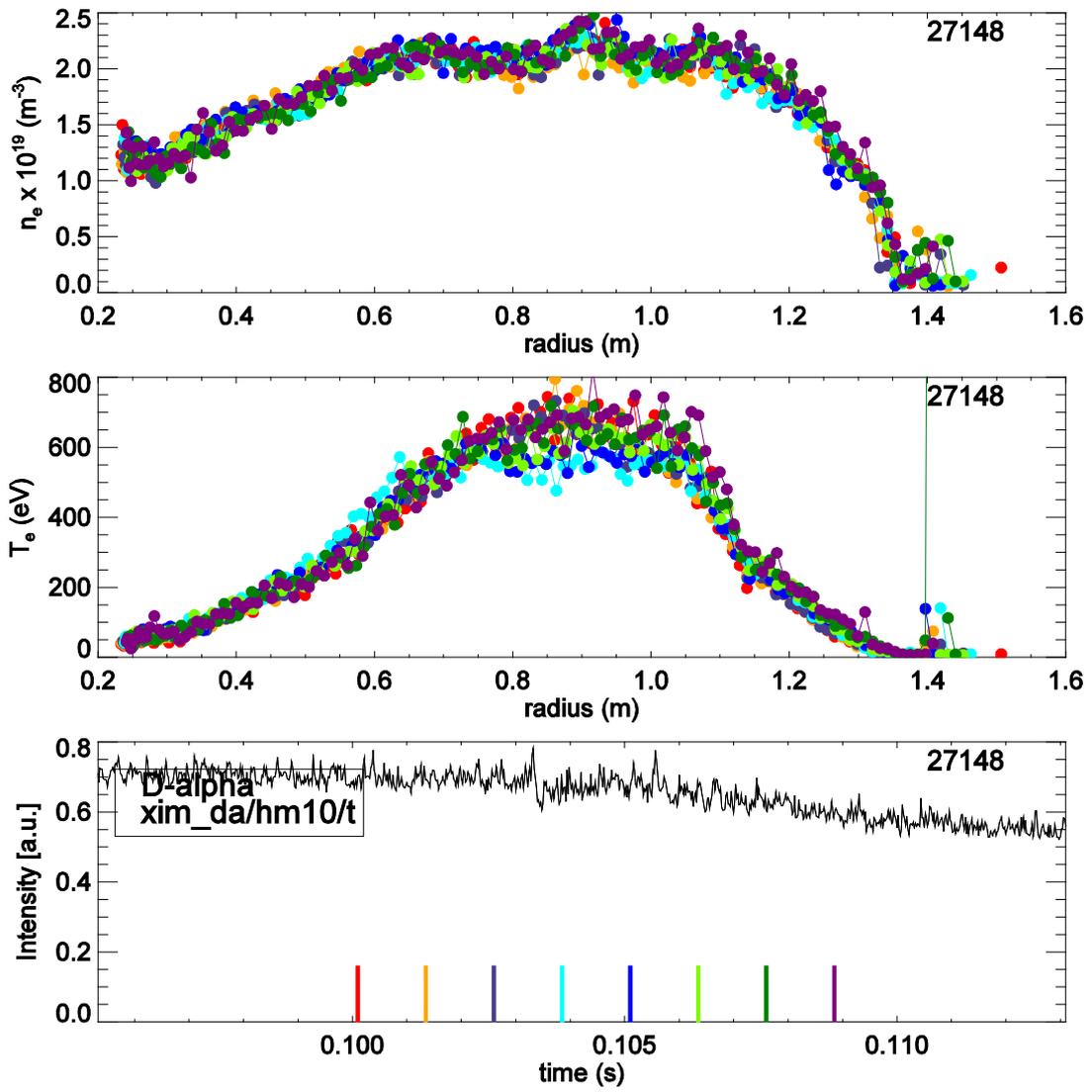

Figure 6.

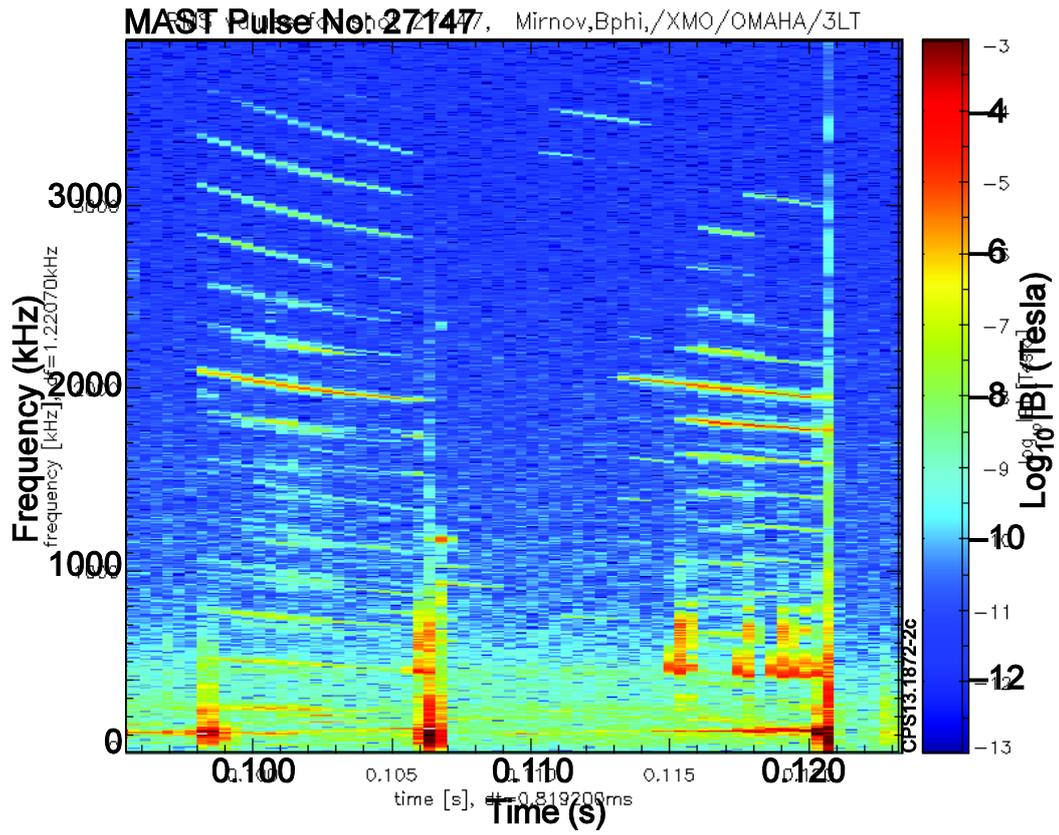

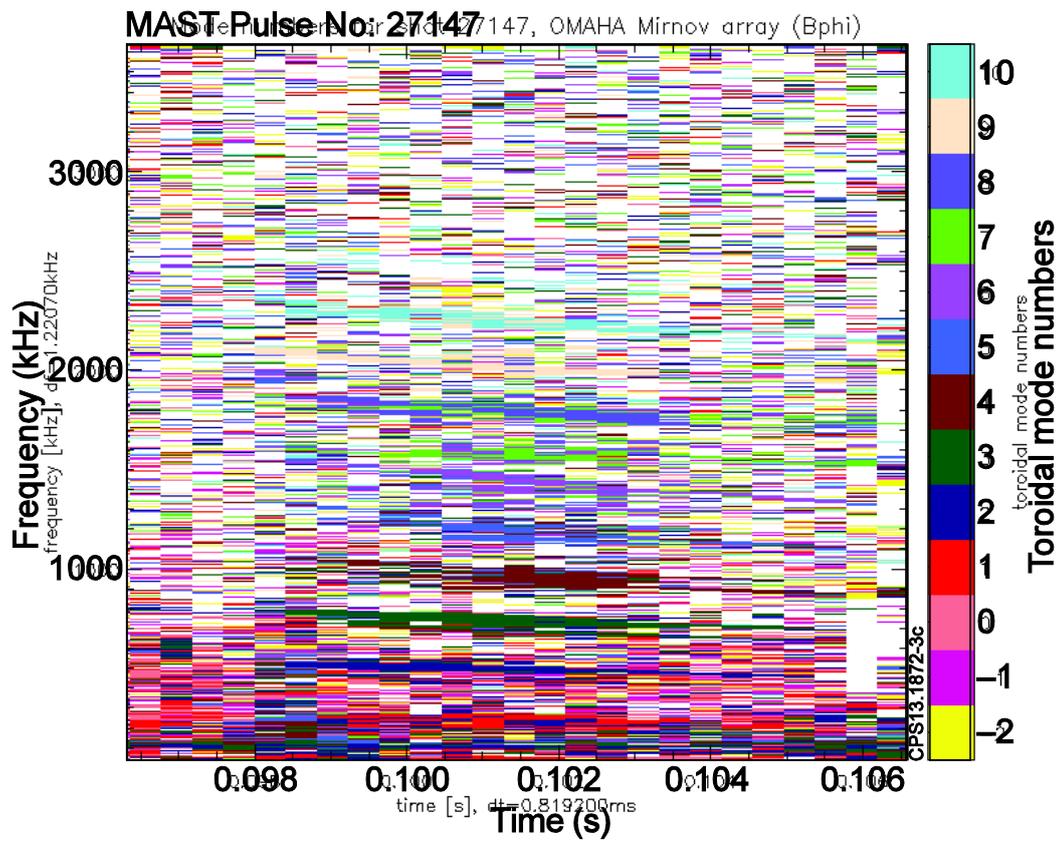

Figure 7.

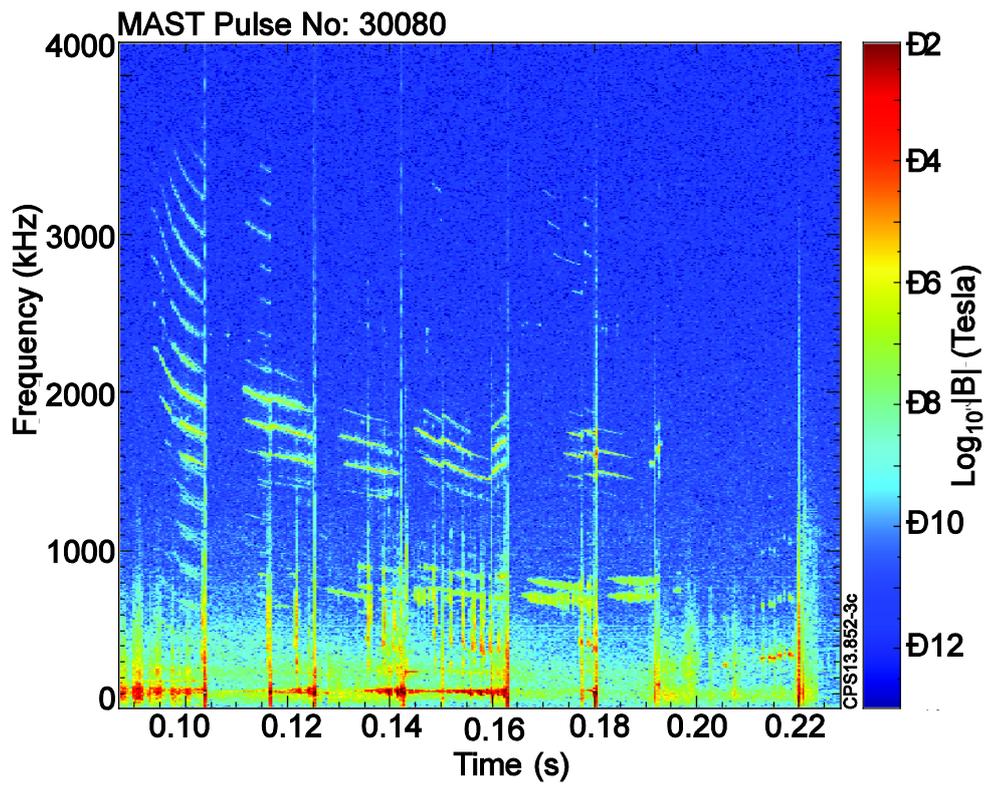

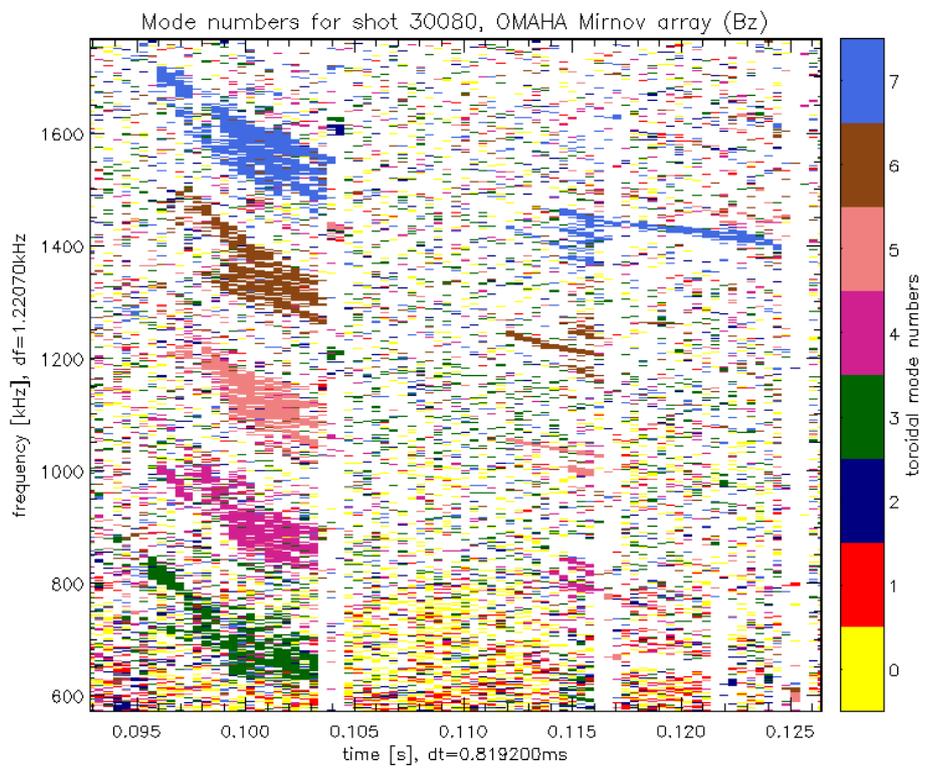

Figure 8.

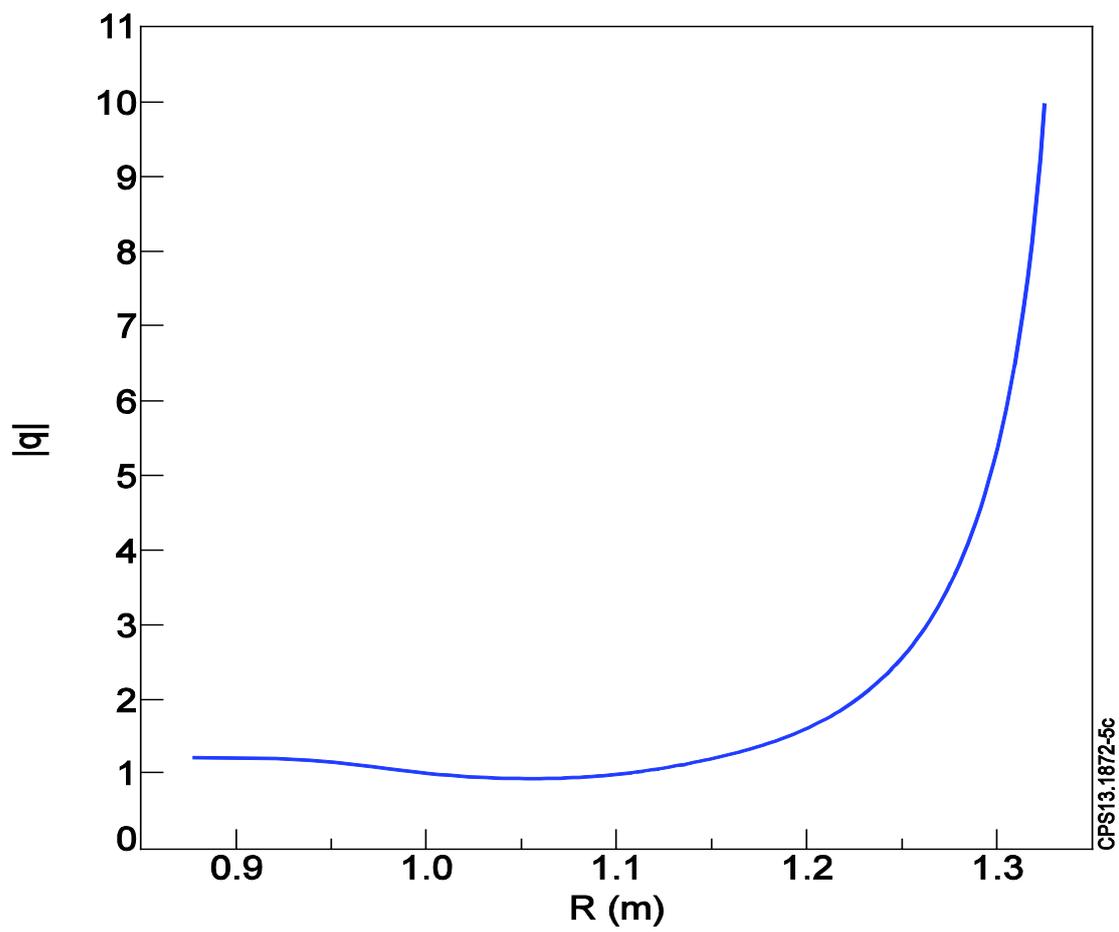

Figure 9.

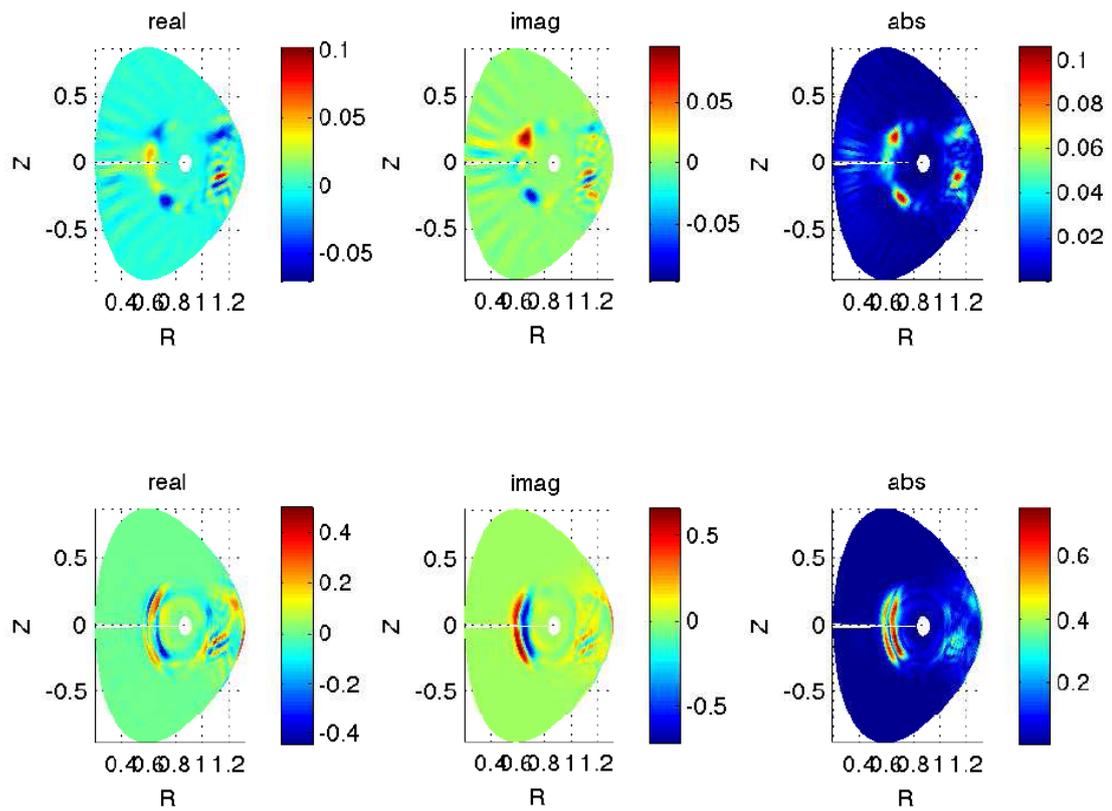

Figure 10.

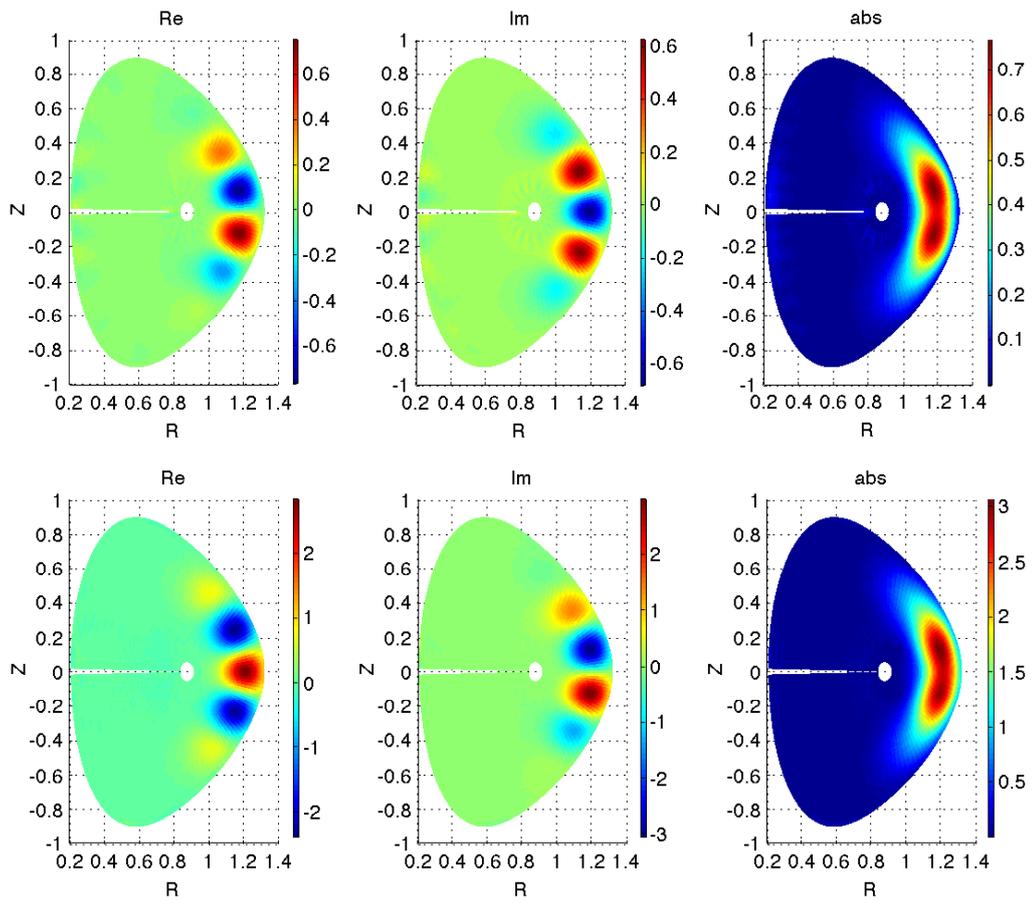

Figure 11.

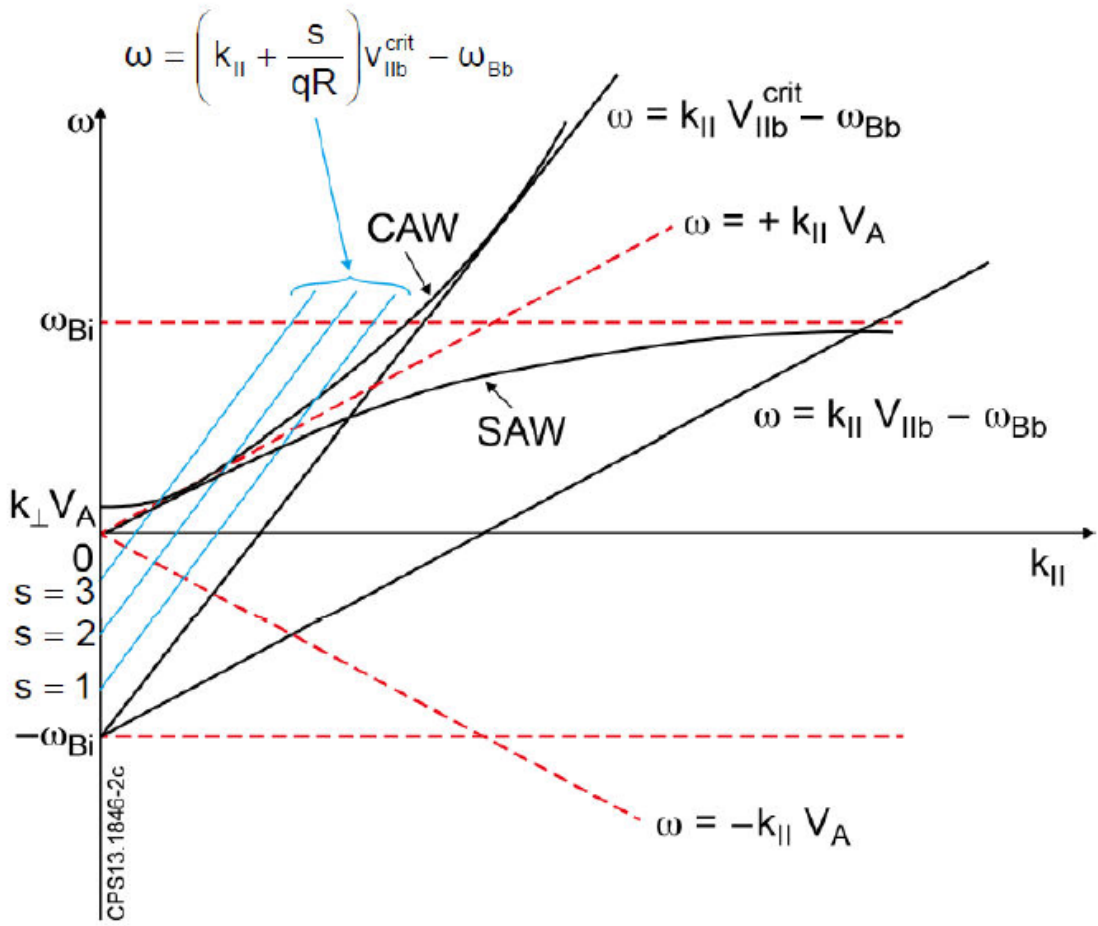

Figure 12.

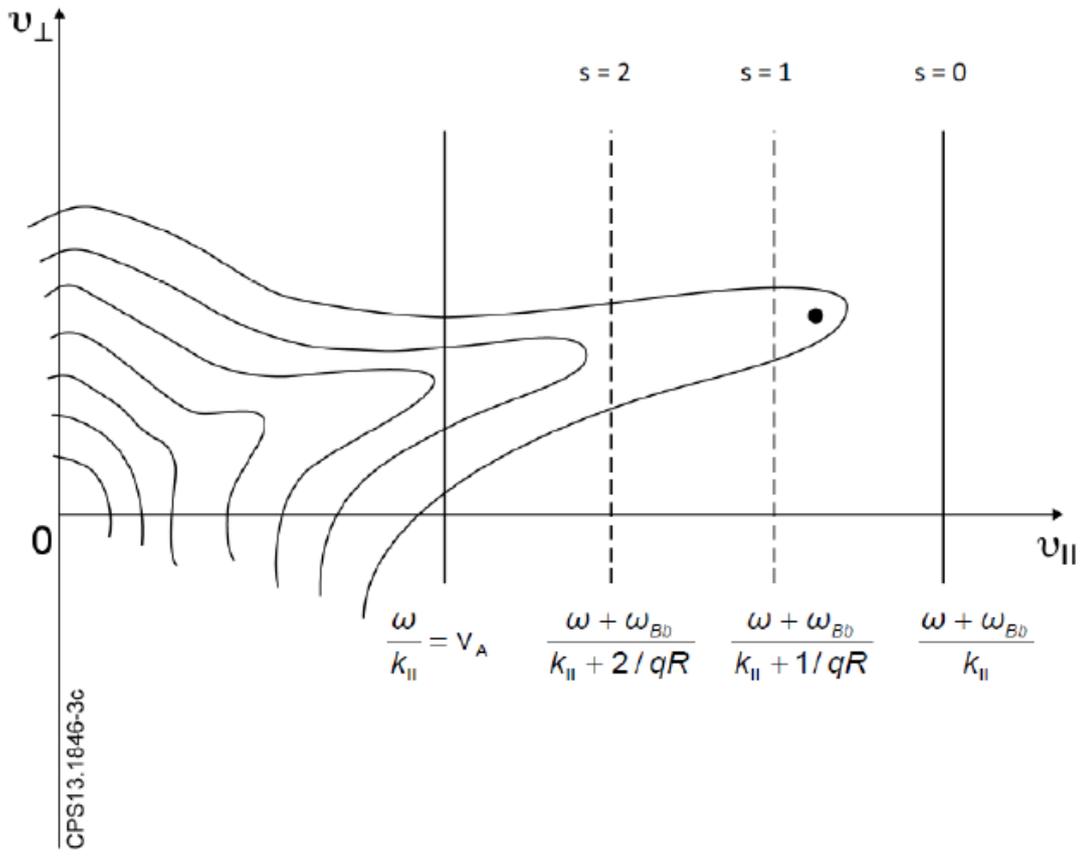

Figure 13.